# Data Privacy and Specimen Pooling: Using an old tool for New Challenges

P. Saha-Chaudhuri[1] and C. R. Weinberg[2]
[1] Department of Epidemiology, Biostatistics and Occupational Health, McGill University
[2] Biostatistics Branch, National Institutes of Environmental Health Sciences, NIH
Corresponding author: paramita.sahachaudhuri@mcgill.ca

**Abstract [250 words]**

Background: In the context of ongoing debate over data confidentiality versus shared use of research data, as raised following the new EU General Data Protection Regulation, we seek to find alternate techniques that can balance these two issues. In particular, we demonstrate that an existing epidemiologic tool, specimen pooling, can be adapted as a privacy-preserving method to enable data analysis while maintaining data confidentiality. Specimen pooling is a cost-effective tool in studying the effect of an expensive-to-measure exposure on a disease outcome, for both unmatched and matched case-control designs. We propose the technique in a new context to analyze confidential data and demonstrate that it can be successfully used to estimate OR of covariates based on aggregate data when individual patient data cannot be shared.

Methods: We demonstrate the application of specimen pooling based on aggregate covariate level and show that aggregated covariate levels can be used in a conditional logistic regression model to estimate individual-level odds ratio parameters of interest. We then show how to adapt the technique as a privacy-preserving method to analyze data from a matched case-control design. A similar approach can be applied for an unmatched design and unconditional logistic regression.

Results: The parameter estimates from the standard conditional logistic regression are compared to those based on aggregated data. The parameter estimates are similar and have similar standard errors and confidence interval coverage.

Conclusions: Pooling can be used effectively to analyze confidential data arising from distributed data networks and will be extremely useful in pharmacoepidemiology.

**Keywords**
Data privacy; conditional logistic regression; specimen pooling; distributed data network;

**Key Messages**
- Data privacy has become a critical issue in research.
- An existing technique of specimen pooling can be adapted to offer a privacy-preserving analytical option for estimating odds ratios from a matched case-control design.

# BACKGROUND
## PRIVACY ISSUES IN RESEARCH
Each year in EU, Canada and elsewhere, regional and federal governments and health care authorities collect an immense amount of personal and sensitive data on health care use, diagnoses, risk factors and behaviors and many other factors. The advent of personalized medicine and the genomic revolution will result in the further gathering of a large quantities of very sensitive data for large segments of the population. Although these data could provide critical information for the management of the health care system, for studying causation of diseases, prognosis and the impact of treatment or prevention efforts, their use is constrained by legitimate concerns about privacy.

While data confidentiality is not entirely a new issue, of late it has attracted attention of many epidemiologists due to the new EU General Data Protection Regulation (GDPR). The previous EU Data Protection Directive was enacted in 1995 with a goal of regulating processing of personal data. However, it has been criticized by the research community as "overly complex, sometimes ambiguous and presenting an obstacle to epidemiological and other research" [1]. Since then, there have been significant advances in digital and mobile technology resulting in data proliferation and related data security issues, both in research [2] and elsewhere [3,4]. Consequently, in 2012, the European Commission proposed to replace the 1995 directive by the GDPR, "with the objectives to harmonize data protection within the EU, facilitate the flow of data across borders and enhance privacy protection." [1] The goal of the proposed GDPR was to strengthen data protection for individuals. In addition to the informed consent principle that has long been a keystone of medical research, in the initial versions of GDPR explicit consent was also sought from individuals so that the data collected during routine healthcare could be used for research projects. Initially some exceptions were allowed for research, but later, the European parliament's Committee on Civil Liberties, Justice and Home Affairs (LIBE committee) proposed amendments to remove most of exceptions, including one for research, and require participant consent for processing health data used in research, with the single exception of "research that serves an 'exceptionally high public interest,' if that research cannot possibly be carried out otherwise."[5]

The proposed amendments triggered intense criticism for their narrow vision of research; the resulting backlash from the research community was evidenced by several articles, all acknowledging privacy concerns, but asking for a better balance between privacy and data use in research, in particular for carrying out epidemiologic studies. [1,5-12] A trialogue between The European Commission, European Parliament and the Council of the European Union ensued, resulting in a final version that was accepted in early 2016, with the goal of enforcement in 2018. The accepted version of the legislation [13] now has a more balanced approach to data protection and research use of data allowing flexibility to "Member States to adapt the rules to fit their existing arrangements and make them more relevant to their own society and culture" [14]

DATA PRIVACY WHILE CONDUCTING EPIDEMIOLOGICAL RESEARCH
Most standard data analysis techniques require individual data (microdata) for estimation and inference; thus, they are not applicable in a setting where microdata cannot be shared without risking disclosure of sensitive information. Frequently, research use data undergo anonymization to remove any identifying information such as name, address, unique identifier (e.g., US Social Security Number). However, it may still be possible to identify participants using combined variables (e.g. if there is a unique person of a certain age, diagnosis, body mass index, town of birth) or auxiliary information available from other sources. [2,3]

Privacy issues have long plagued other fields as well, such as survey research, resulting in the development of analytical approaches that can extract valuable information from microdata while protecting data privacy. Privacy-preserving statistical analyses strive to maintain a balance between data use and confidentiality. Privacy-preserving methods have many uses, in particular for analysis of sensitive socio-economic, financial, infectious disease and genetic/genomic data. Recently, agencies generating and using distributed data such as disease registries or healthcare surveys where data are kept secured at regional level and not shared between regions (see Fig 2 for a schematic of distributed data), have been looking to adopt privacy-preserving analyses to avoid identification of participants. Data privacy is also relevant for epidemiological research, especially in the light of the new GDPR, as epidemiologic studies use secondary data analysis or reuse data for research that were collected for other purposes, e.g., routine healthcare delivery, and requiring an informed consent for a particular research project may not be possible or may result in highly non-representative data.

GOAL OF PRESENT MANUSCRIPT
We agree that there needs to be open and fair data access for responsible research in such a manner that data privacy is maintained. With that objective in mind and without going into regulatory and legal aspects, the goal of the present manuscript is to demonstrate that existing analytical tools may be adapted for epidemiologic research, such that the important scientific questions can be answered while respecting confidentiality of individuals[15]. In particular, we show how the approach of specimen pooling can be adapted to analyze confidential data from a matched case-control study.

BACKGROUND ON SPECIMEN POOLING AND MICROAGGREGATION AS PRIVACY-PRESERVING METHOD
The idea of specimen pooling originated during World War II in a slightly different context when blood from military recruits were pooled to allow efficient identification of recruits with syphilis[16]. Pooling of biospecimens is used predominantly in infectious disease setting where multiple specimens are pooled together to identify the infection status of the pool of samples, because if all of the contributing samples are negative for an infectious disease (e.g., HIV), the pooled specimen will also be negative whereas (with a sufficiently sensitive assay employed) the pooled specimen will be positive otherwise. Such a strategy leads to a reduction in the overall assay cost. The method has also been recommended for estimating the prevalence of HIV in a population, while concealing the identities of individuals who are HIV-positive[17]. We focus on specimen pooling for estimating an exposure OR using a logistic regression model for a binary disease outcome.[18, 19] We demonstrate that this strategy extends easily to provide a privacy-preserving analytical method[15] for analysis of matched case-control design or a matched design with a binary endpoint. The idea of specimen pooling in this context is similar to a privacy-preserving technique called micro-aggregation that has been used frequently for estimating parameters of a linear regression model.[15] The rest of the manuscript is organized as follows. In the methods section, we outline our specimen pooling approach for matched case-control study and show how it can be adapted as a privacy-preserving method. We provide simulation results and a real data example in the results section and conclude with a discussion of the strengths and limitations of the proposed approach.

**METHODS**
In this section, we first describe pooled logistic regression for a matched design and follow up with modifications for distributed data. Since the analysis is based on aggregate rather than individual-level data (microdata), even when the aggregated covariates pass through firewalls, patients' identities are always protected. We demonstrate the approach for a binary outcome when the data are collected using a matched case-control design.

NOTATION AND LOGISTIC MODEL FOR MATCHED CASE-CONTROL DATA
Let $U$ denote the level of a continuous exposure of interest and $D$ (1 for cases and 0 for controls) denote the disease status. To simplify the exposition we first consider a pair-matched case-control study. Assume the following logistic model for the $j^{th}$ subject ($j=1, 2$) of $i^{th}$ of N matched pairs:

$$logit \Pr(D_{ij} = 1 | U_{i,j} = u) = \alpha_i + \beta u, \ j = 1,2; i = 1, 2, \ldots, N. \quad (1)$$

where $\beta$ denotes the log odds ratio (OR) associated with unit increase in exposure and $\alpha_i$ denotes the pair-specific baseline log odds of disease and is considered a nuisance parameter. Denoting the case exposure in the $i^{th}$ pair as $u_{1,i}$ and the control exposure as $u_{0,i}$, the contribution of the $i^{th}$ matched pair to the conditional logistic regression likelihood is:

$$\frac{e^{\beta u_{1,i}}}{e^{\beta u_{1,i}} + e^{\beta u_{0,i}}}$$

Additional covariates and effect modifiers can be accommodated in the usual manner. This likelihood formulation extends readily to matched sets with more than one control per case. Any standard statistical software can be used to obtain a maximum likelihood estimate of $\beta$ and its associated standard error. In particular, one can use the `clogit` function in the programming language R for fitting a conditional logistic regression model from a matched case control design. Additional confounders and effect modifiers can be accommodated easily. However, effect of a matched-pair-specific covariate is aliased with the baseline log odds and hence cannot be estimated using a matched design.

POOLED ANALYSIS OF MATCHED CASE-CONTROL DATA
To allow estimation of $\beta$ based on pooled exposure level for data from a matched case-control design, we follow a two-stage procedure. In the design stage, pools are formed and exposure is measured in the pools. In the second stage, the pooled levels are used in the regression model to estimate $\beta$.

Stage 1: Pooling Strategy
Let us consider the 1:1 matched design as before and let $N=gk$, where $N$ denotes the number of matched pairs and $g$ is the intended poolsize. In the design stage, we randomly partition $N$ matched pairs into $k$ groups, each of size $g$. For each such set of pairs, equal-volume aliquots from the $g$ case specimens are combined to form a single pooled case specimen, and the corresponding $g$ control specimens are pooled to form a single pooled control specimen. Unlike an unmatched case-control design where control pools are formed independently of the case pools[18], in the matched design, case and control pools are formed in the same manner so as to keep the connection between the matched sets.

Stage 2: Pooled Model and Analysis
Once the matched sets are grouped, instead of $N$ matched sets, there are now $k$ matched sets and $2k$ pools (each with $g$ matched sets). If we index the $g$ exposures (or specimens) in a typical pooling set by $i$, the measured exposure in that pooled specimen is $\overline{U^j} = \frac{1}{g}\sum_{i=1}^{g} U_i^j$ where $U_i^1$ denotes the level for the $i^{th}$ specimen for the cases and $U_i^0$ denotes the $i^{th}$ specimen for the controls. We also pool the covariate data, either by pooling biospecimens as for the exposure of interest, or by taking averages for values corresponding to each case or control pooling set. In other words, rather than measuring exposure for each individual in the study, only the aggregate exposure level is measured for the pool, thereby giving rise to *k* matched sets of pooled covariate measurements.[19]

A logistic regression model, such as Eq. (1) induces a logistic model for the pooling sets[18]. For a matched case-control design, the contribution of the $j^{th}$ matched pool to the conditional logistic regression likelihood will be [19]:

$$\frac{e^{\beta v_{1,j}}}{e^{\beta v_{1,j}} + e^{\beta v_{0,j}}}$$

where $v_{1,j}$ denotes the pooled (aggregate or sum of) exposure level from the cases and $v_{0,j}$ denotes the pooled exposure level from the controls from the same matched pool. As before, standard statistical software can be employed to analyze data from a pooled and matched case control design. Additional confounders/covariates that have been aggregated, as described above, can be accommodated easily. Estimation efficiency for the primary association can be improved by utilizing available subject-specific information at the pool allocation stage[20].

Inclusion of an effect modifier (EM) requires pools to be formed by both outcome and EM status. For example, if gender (male vs. female) and age (young vs. old) are of interest as potential EMs, then pooling groups have to be formed separately for young males, young females, old males and old females. Consequently, only categorical effect modifiers that require a limited number of categories can be readily included in the model.

CONFIDENTIAL AND DISTRIBUTED DATA

We now demonstrate how ideas of specimen pooling can be adapted to analyze distributed data in such a way that data privacy is maintained. To formalize the idea, we consider a simple horizontally partitioned data network with multiple nodes and one analytical center. A horizontally partitioned datasets splits the data into several subsets where each node or center only owns complete records of a subset of participants. This is in contrast to a vertically partitioned dataset where each node owns partial covariate information of all subjects. A schematic of horizontally and vertically partitioned data can be found in Figure 1.

[Figure 1 here]

For such a horizontally partitioned data, each node holds a subset of records and cannot share their data with other nodes. An analytical center (AC) can be selected that has contact with the nodes and can instruct and supervise statistical analysis. In addition, the center will generally have the capability of analyzing data; however, due to privacy concerns, the nodes may not be able to share microdata with the AC. A designated node can act as the AC, or representatives from nodes may constitute an AC, or a suitable third party can serve as the AC. We also assume that while individual information cannot be shared between nodes or with the AC, aggregate level covariate information such as the total counts for binary or categorical variables and sum/average for continuous variables can be shared between nodes and centers. We assume that nodes do not collude with each other and use additional measures such as secure summation[21] in addition to aggregate covariate information to further protect microdata. Finally, the center can direct the nodes as to how to aggregate the covariates and can receive aggregate data from nodes without compromising data privacy. A schematic of distributed data with four nodes and one AC is shown in Figure 2.

[Figure 2 here]

Exposure pooling for confidential data will follow a protocol similar to that of the standard pooled analysis. Once the microdata from a matched case-control design are available at node level and the total number of matched sets, *N,* is computed, the first task is deciding the poolsize *g*, followed by forming pools according to this poolsize and finally, aggregating covariates according to the strategy. The second step is to use aggregate covariates to estimate the log ORs associated with the covariates as outlined above.

While multiple different values of *g* can be accommodated in a pooling design[19], for simplicity, we recommend using at most two different values of *g* (e.g., *g* = 5 and 3) such that all individual data are used for analysis. For example, if *N=2389* with *1:m* (*m>=1*), matching, one can use *476* pools with *5* matched sets each and *3* pools with *3* matched sets each, so that all subjects are included in the pooled analysis. If only one value of *g* is used, a small number of observations are excluded, e.g. with the same *N=2389*, *477* pools with *5* matched sets each can be formed and used in the pooled analysis. The remaining 4 matched sets would need to be excluded from the pooled analysis. Such a strategy offers a judicious balance between design complexity and maximal data use. Use of multiple matching ratios in a design (e.g., 1:1 matching and 1:2 matching) will also impact choice of poolsize as the pooling strategy would have to group together only matched sets of similar structure.

Once *g* is decided, pooling groups should be formed at random simply by partitioning *N* matched sets in *k* groups of size *g* each (assuming *N=k.g*). Covariates are then aggregated by summing the levels for all members in the pool. The aggregated covariate levels are recorded for each of these *k* pools separately for cases and their matched control. These aggregate values now can be shared without compromising privacy of the microdata. In the final step, the aggregate values are used in the pooled conditional logistic regression model to estimate targeted log ORs.

In a distributed data setting, an additional complexity arises due to the horizontal partitioning of the data, in that each node typically owns a subset of data and cannot share the data with external nodes or AC without potentially compromising privacy. Aggregating covariates in such a scenario typically involves secure summation [21]. We outline the pooling strategy step by step below and share a toy example in the next section.

[Table 1 here]

1. After applying appropriate inclusion and exclusion criteria, each node sends the number of matched sets to the AC to determine the total number of matched sets included in the study cohort. Let *N* be the number of matched sets.

2. The AC determines the poolsize(s) *g* to be used (from the rest of the steps we approximate *N=k g* for simplicity of disposition).
3. The AC creates the pooling sets by randomly partitioning the matched sets into *k* groups, with the $i^{th}$ pool consisting of matched sets $(N^i_1, i_1), (N^i_2, i_2), \ldots, (N^i_g, i_g)$, where $N^i_j$ denotes the id of the node and $i_j$ denotes a particular matched set (see Table 1).
4. The AC sends a table similar to Table 1 to nodes to help identify pools and help aggregate levels of the variables. Nodes work among themselves to aggregate appropriate covariate values according to the matched id datasets and then pass the aggregate values to the AC. Privacy-preserving techniques, such as secure summation, can be employed here.
5. The AC uses a dataset as in Table 2 and runs conditional logistic regression using matched pools (case pool or matched control pool) as outcome and pooled covariate values. In Table 2, we demonstrate such data for a 1:1 matched design and two variables V1 and V2.

[Table 2 here]

ACCOMMODATING CONFOUNDERS, EFFECT MODIFIERS AND TRANSFORMATIONS FOR DISTRIBUTED DATA

In addition to a primary exposure, confounders can be accommodated simply by aggregating the levels according to the pools. However note that for a matched case-control design, a stratum-specific confounder that varies between matched strata, but not within strata, is aliased with the random effect $\alpha_i$. Consequently, the effect of such a confounder on risk of the outcome cannot be estimated. However, the associated confounding has been controlled by the matching. A similar limitation also applies to a pooled model in that the OR of stratum-specific confounders cannot be estimated. They can, however, be included as effect measure modifiers.

In a traditional epidemiologic setting where aggregate exposure is measured in the pools, only *categorical* effect modifiers can be accommodated, which is done by requiring that pooling be done conditional on the levels of *effect modifiers (EMs)*. Moreover, pooling within the levels of an EM renders the main effect of the EM aliased with the random effect $\alpha_i$ and hence it is not possible to estimate the main effect of the EM. Finally, if multiple EMs are to be included in the model, the advantage of pooling can be reduced due to sparsity, as some combination of EMs may not have enough subjects (g or more) to support pooling.

In contrast, the above limitations disappear for distributed data setting simply because individual covariates are available at the nodes. Consequently, we do not require pools to be formed conditional on the levels of both the effect modifiers and the outcome. Simply conditioning on the outcome status will allow us to treat the terms involving an effect modifier as confounder terms and enable us to estimate ORs that involve EMs. Similar to a stratum-specific confounder, one cannot estimate the main effect of a stratum-specific EM, but the main effect of a subject-specific effect modifier can be estimated using the proposed approach for distributed data.

Transformations can also be handled easily. For example, if log transformation is used for one of the covariates $X$, instead of sending $\sum x_i$, pooled levels of the transformed variable $\sum \log x_i$ can be shared for modeling purposes. Derived variables, such as body mass index and creatinine-corrected urinary levels, can also be accommodated by calculating the index prior to summing across individuals. Care should be taken in the choice of *g* when multiple functional forms of the same covariates are included in the model. For example, if a cubic power and all lower powers of a covariate $X$ are included in the model, then use of $g \leq 3$ would allow one to algebraically infer the individual covariate values. However, even in this case, it would not be possible to know which individual participants had which values.

INFERENCE FOR PARAMETERS AND MODEL SELECTION

In a distributed data setting with a matched design, it is possible to assess association of the exposure with the outcome using pooled exposure determinations. The pooled model is simply a conditional logistic regression model with pooled or aggregate covariate levels instead of individual level covariates. Hence, in addition to the parameter estimate, estimated standard error (SE), confidence interval and associated p-value can be obtained for the exposure and other variables (including transformations), except for stratum-specific variables. Consequently, model selection involving confounders, different transformations of the variables, and EMs will be possible. Two nested and competing models, fit using pooled data, can be compared using a likelihood ratio test or AIC. Since the microdata is available at the node level for distributed data, akin to bootstrap procedure, the step of formation of pooling groups and parameter estimation can be repeated several times to obtain several parameter estimates that can capture the breadth of the distribution of the estimates.

**RESULTS**

TOY EXAMPLE

[Figure 3 here]

We first demonstrate our approach for distributed data using the toy example from Figure 3. We consider three nodes (blue, red, green) each with three matched sets. Each matched set consists of one case and two controls. The covariates are: age (A; continuous in years), gender (G; male or female) and a fictional biomarker (M; continuous).

[Table 3 here]

The simulated individual level data for all 27 participants from 9 matched strata are shown in Table 3 and we would like to fit the following model:

$$logit \Pr(D_{ij} = 1 | A_{ij} = a, G_{ij} = g, M_{ij} = m) = \alpha_i + \beta \times \log a + \gamma \times g + \delta \times m + \vartheta \times \log a \times m.$$

When individual data is available, the above model can be fit easily using any statistical software package.

For pooled modeling, the first stage is to identify an appropriate pool size $g$ and then randomly partition within the strata the matched sets into sets of size $g$ to create the pooling groups. Here we use $g=3$ giving rise to 3 sets of pools, each with one case pool and 2 control pools. A random partition of the 9 matched strata is depicted in Figure 3 by the rectangles, where the first case pool consists of one case each from the three nodes. The next case pool is formed by one case from blue node and two cases from the red node. The final case pool is formed by one case from blue node and two cases from the green node. For each case pool, their corresponding controls comprise the two matched control pools giving rise to 9 pooled sets (three case pools and their matched (six) control pools, similar to the original 1:2 matching scheme).

[Table 4 here]

The variables are aggregated according to the pooling strategy keeping in mind the requisite transformations in the intended logistic regression model. Noting that instead of aggregating age and then taking the logarithm, the correct order is to take the logarithm and then aggregate the log(age). Similarly, for the interaction between log age and the marker, we first multiply log(age) with marker and then pool according to the pooling scheme. The pooled data is shown in Table 4. Afterwards, any statistical software that can analyze data from matched case-control study can be used to analyze these data by treating the pools as units of analysis and treating pooled levels as the covariate levels. Along with the estimated log OR based on $\beta, \gamma, \delta, \vartheta$, SEs can be obtained from the output of the software. In order to test for the confounder coefficient, $\delta$, or, for the interaction parameter, $\vartheta$, one can use a t-test or likelihood ratio test. A likelihood ratio test can also be used to simultaneously test for multiple parameters, for example, an effect of age categories.

SIMULATION EXAMPLE

The structure of our simulation is motivated by a recent study of incretin-based drugs and congestive heart failure (CHF), which used a matched case-control design[22]. In the study, the authors used data from United Kingdom's Clinical Practice Research Datalink [23] and matched 1118 cases to ≤ 20 controls per case for a total of 17,626 controls. Matching factors were age, duration of treated diabetes, calendar year, and time since cohort entry. For simplicity, we consider a matched case-control design with 1020 matched sets, each with a fixed matching ratio of 10 controls per case and considered 500 such datasets for each parameter combination.

The primary exposure U was log normally distributed and was assumed to be a confidential variable. The secondary exposure X was binary with a prevalence of 0.4 and independent of U. A confounder $Z_1$ was distributed as normal with a correlation of 0.35 with log(U). A continuous EM $Z_2$ was generated as standard normal independent of U, X and $Z_1$. The stratum-specific intercept for the matched sets was generated as normal with mean of -3.0 and SD of 2. Finally, the following model was considered for generating the binary outcome:

$$logit \Pr(D_{ij} = 1 | U, X, Z_1, Z_2) = \alpha_i + \beta\, U + \gamma\, X + \delta\, Z_1 + \omega\, Z_2 + \vartheta\, U\, Z_2.$$

For pooled analysis, we considered pools of sizes: g = 4, 6, 10 conditional on the outcome status, leading to 225, 170 and 102 matched clusters (and matched case pools). Unlike traditional epidemiologic setting where presence of a categorical EM would require pooling to be done conditional in addition on the EM status, for distributed data, we treat each additional term involving the EM as confounders, allowing us to accommodate continuous EM and to estimate the main effect of the EM. All simulations were performed in the programming language R.

We considered several combination of parameter values and show the results for the following set: $\beta$ = 0.3 (corresponds to OR = 1.35), $\gamma$ = 0.2 (OR= 1.22), $\delta$ = 0.15 (OR= 1.16), $\omega$ = 0.09 (OR= 1.09) and $\vartheta$ = 0.05 (OR= 1.05). In Figure 4, we plotted the parameter from the standard analysis and pooled analysis (with g = 6) estimates for all 500 simulations to assess the agreement between the two sets of estimates. For reference, we added a diagonal line (dashed) and lines corresponding to the true parameter values (dotted vertical and horizontal lines). We also added a lowess line (solid, red) for predicting the pooled estimate based on the unpooled estimate, to compare the estimates. The better the agreement between the pooled estimate with the unpooled estimate, the closer the lowess line to the diagonal line, although some degree of regression to the mean is to be expected, especially when larger poolsize is used. In Table 5 we summarize the simulations to report the average parameter estimate, Monte Carlo SE (MCSE), average model-based standard error (ModelSE) and coverage probability out of 500 simulations for unpooled (standard conditional logistic regression) and pooled conditional logistic regression with $g$ = 4, 6, 10, where MCSE is defined as the standard deviation of the parameter estimates over the 500 simulations and ModelSE is the average of the 500 estimated standard errors from the conditional logistic regression.

As can be seen from Figure 4, the red lowess line generally coincides with the diagonal line indicating that generally, the pooled and unpooled estimates demonstrated good agreement for all parameters of the model we considered. In Table 5, comparing parameter estimates from the unpooled analysis with pooled analysis, we see that in general, the pooled analysis performs comparably to the unpooled analysis, with a tendency to show slight bias away from the null, especially with increasing pool size, presumably due to reduction in effective sample size. The model-based SE for pooled analysis was also slightly inflated as compared to model-based SE for unpooled analysis. However, the coverage for the 95% confidence interval for pool sizes 4 and 6 was in line with the nominal level, although slightly inflated for pool size 10.

REAL DATA EXAMPLE
We demonstrate our method using [one or two] real datasets. We reanalyzed data from a matched case-control study of infertility after induced and spontaneous abortion.[24] In the original study, 83 women with secondary infertility were each matched to 2 controls on age, parity and education level. We used the study data as available from CRAN (R data infert()). For the purpose of demonstration, we made some modifications to the dataset. First, we excluded one matched set with only one control, so that all 82 matched strata under consideration had exactly two controls matched to a single case (data0). Since pooling reduced the effective sample size, we bootstrapped strata to create a pseudo-datasets with 1000 matched sets, each with exactly two controls matched to a single case (data.unpooled). For pooled analysis, we used *g=2* and grouped together two sets at random to pool covariates (data.pooled). For each of these three datasets, we modeled the outcome (secondary infertility - yes/no) with the analysis based in turn on two variables (any induced abortion - yes/no and any spontaneous abortion - yes/no) and their interaction and used conditional logistic regression to estimate the ORs associated with the variables and interaction. All analyses were done in the programming language R. In Table 6, we report the ORs and associated 95% confidence interval (CI) for each of the three datasets.

     The OR estimates based on the pooled conditional logistic regression model is in line with the unpooled estimate from the original data (data.0) and the bootstrapped version of the data (data.unpooled). The 95% CIs from the pooled model were similar to those computed using the model based on unpooled (bootstrapped) data, but slightly shifted. Overall the performance of the pooled approach was comparable to the unpooled approach.

# DISCUSSION

## SUMMARY OF PROBLEM AND OUR PROPOSAL

In this manuscript, we introduce a new application of specimen pooling as a privacy-preserving analytical technique for analyzing confidential data. Large epidemiologic studies often combine data from various sources to study associations between covariates and an outcome of interest. For example, to examine post-marketing safety signal of marketed drugs, multiple registries may be combined. Large datasets are needed to assess associations of small magnitude. However, although ideal, it is often not possible to combine all the data into a single file due to privacy restrictions that prohibit data sharing beyond the owners of the registries. In such a situation, aggregate data can be used to assess exposure-outcome association without revealing individual level covariate combination.

## STRENGTHS AND LIMITATIONS

The pooling technique allows consistent estimation of effects associated with a primary exposure and confounders. In addition, effects of an effect modifier can be assessed by recognizing that the terms involving EM can be treated as an additional confounder. Furthermore, transformation of variables, such as log or polynomials, can be accommodated and in each case, the relevant parameter(s) can be tested using a Wald t-test or likelihood ratio test, allowing researchers to perform model selection. To summarize, exploratory modeling is possible with pooled data and does not require any novel tool. Moreover, standard statistical software can be used for analysis and unlike existing approaches [25, 26] does not impose a heavy computational burden on nodes that may have limited statistical support.

The approach is not without its limitations. It is important to keep track of the pooling strategy as well. One important limitation of aggregation is that for non-normally distributed covariates, the distribution of aggregate values is not the same as the distribution of the individual values. Consequently, summary statistics such as the median or quantiles of the individual level covariate cannot be extracted from the aggregate data. However, frequently, nodes can share summary tables allowing researchers to characterize the distribution of the covariate subject to pooling, albeit crudely due to privacy restriction. In spite of the limitation restricting one's ability to characterize the marginal distribution of the covariate after pooling, the association between the covariate and outcome is preserved even when aggregation is used. It is the latter feature that allows us to estimate the individual-level OR using pooled covariate data.

Our proposal is limited in scope and does not address all the legal and confidentiality issues. However, our hope is that while we continue to seek fair data access policy for research, pooling and other privacy-preserving techniques already in use in the statistics and computer science literature can and should be adapted for epidemiological research.

## POOLSIZE RECOMMENDATION

An important detail for specimen pooling or micro-aggregation is the choice of the pool size. If small pool sizes are used, then there is a risk of disclosure of confidential participant information. However, as shown in our simulations, larger pool sizes tend to bias estimation. As a result, we recommend pool sizes between 3-6 and no more than 10.

## FUTURE DIRECTIONS

We show here that the approach of specimen pooling, which has a history of use in epidemiologic contexts as a cost-efficient approach to study an exposure that is expensive to assay in relation to presence or absence of a disease, can be adapted successfully as a privacy-preserving analytical tool for analysis of confidential and/or distributed epidemiological data. We propose a pooling approach that uses only aggregated covariate levels, thereby making it impossible to link individual data to individual participants. This general strategy can be used in conjunction with data from either a matched or an unmatched case-control study design. In this context, pooling is equivalent to micro-aggregation, which has been used extensively as a statistical disclosure limitation technique for linear models for confidential financial and/or survey data, among others[27-29]. However, micro-aggregation has not previously been proposed for logistic regression. We demonstrated the feasibility of this technique for analysis of confidential data using toy example, a simulation example and a real data example. Other pooling techniques could also be adapted for analysis of confidential data, in fact one of the primary use of specimen pooling is in infectious disease setting where it has helped to protect patient privacy. This will give a new direction to epidemiologic research involving distributed, confidential datasets.

**Tables and Figures**

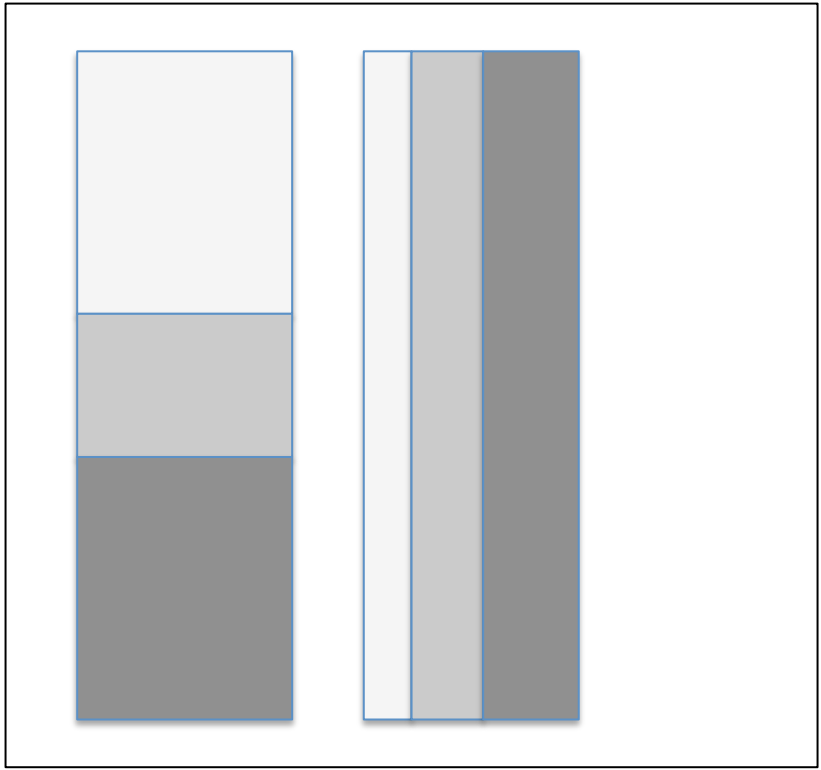

**Figure 1:** Models of Horizontally and vertically partitioned data as adapted from Karr, et al. (2007). Each row is for one subject, and columns are for covariates. (a) Horizontally partitioned data. Data subjects are partitioned among database owners or nodes. (b) Vertically partitioned data. Covariates are partitioned among nodes.

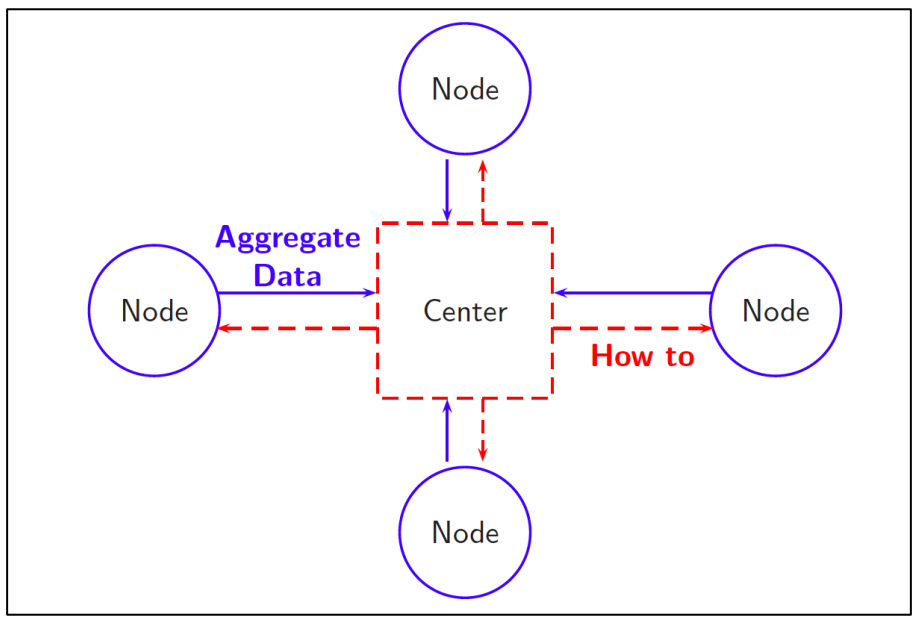

**Figure 2:** Schematic of distributed data.

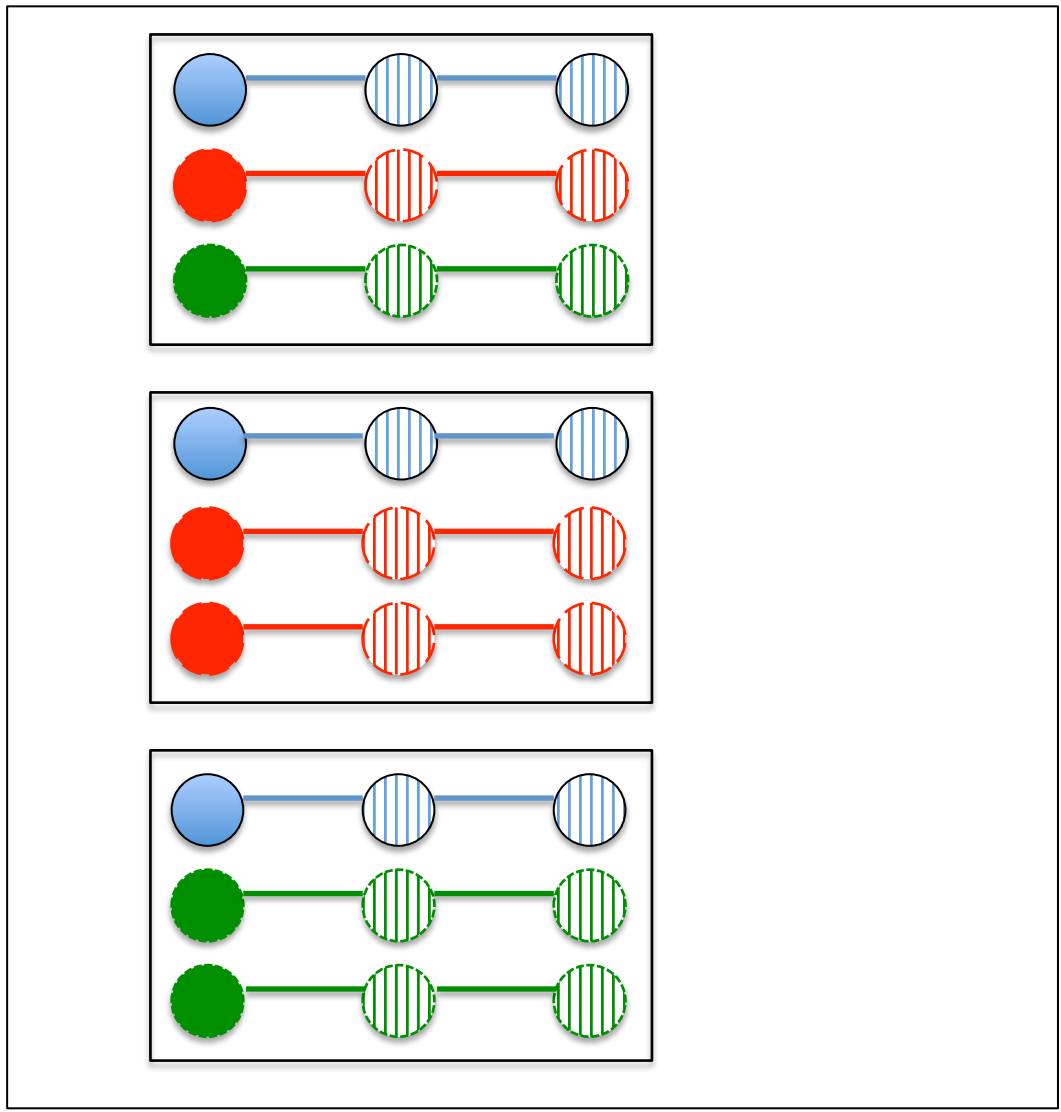

**Figure 3:** Schematic of pooling strategy for a *1:2* matched case-control study with three nodes (blue, red and green) each with three matched sets. Filled circles represent the cases, patterned circles represent the controls, and a line joins the case and controls in each matched stratum. Boxes surround the pools. A poolsize of *g=3* is used. The first pool is formed by combining three matched sets, one from each node. The second pool is formed by combining one matched set from the blue node and two matched sets from the red node. Third pool is formed by combining one matched set from the blue node and two matched sets from the green node.

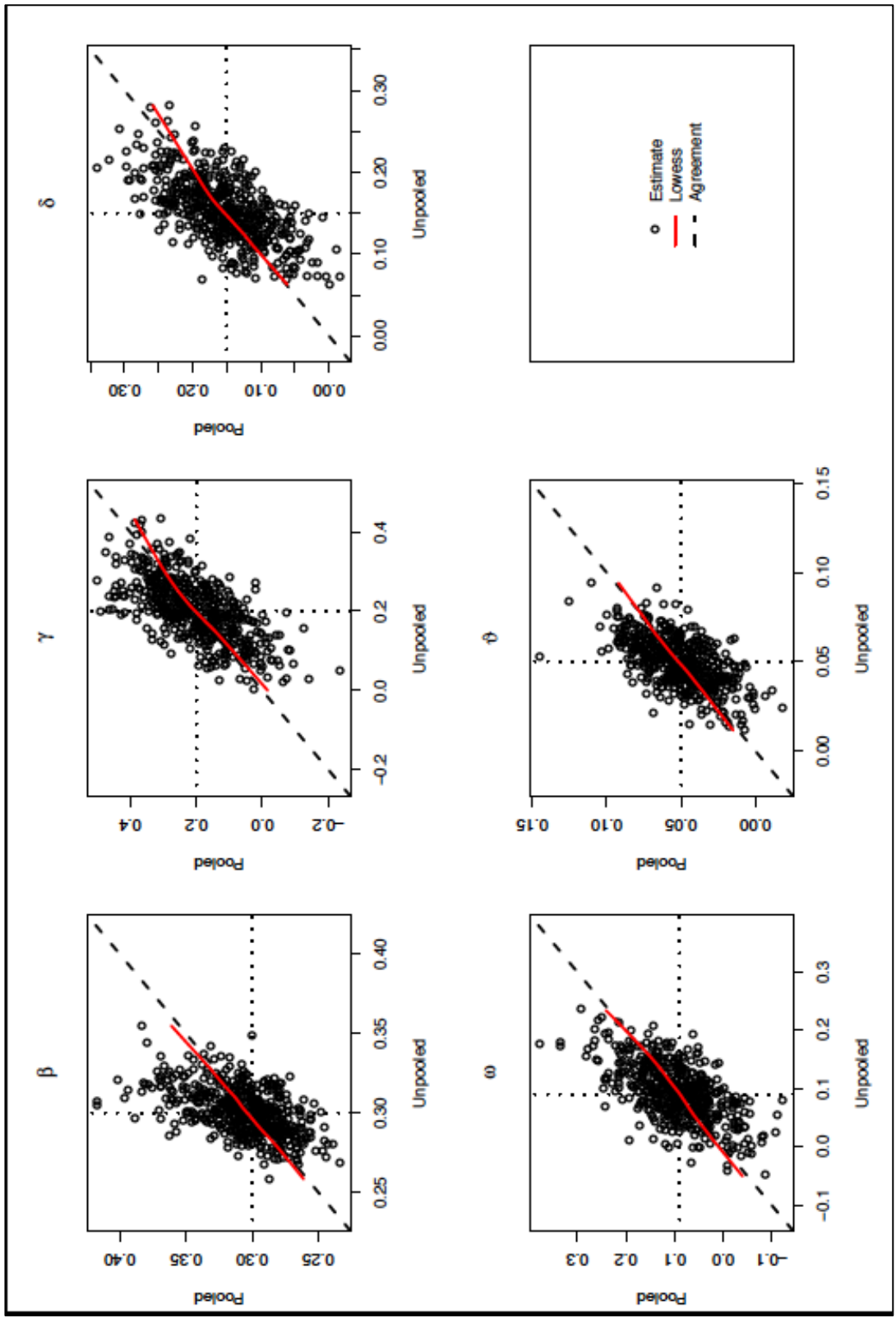

**Figure 4:** Comparison between unpooled and pooled estimate ($g = 6$).

**Table 1:** Id dataset will consists of the pool id, node id and case/control ids contributing to a specific pool. For example, the i$^{th}$ pool consists of subject (case or control) $i_1$ from node $N^i_1$, subject $i_2$ from node $N^i_2$, ..., subject $i_g$ from node $N^i_g$. Multiple subjects from a single node may be present in a given pool, but each subject is present in one and only one pool.

| Pool id | Node id, matched set id | | | |
|---|---|---|---|---|
| 1 | $N^1_1, 1_1$ | $N^1_2, 1_2$ | ... | $N^1_{g}, 1_g$ |
| ⋮ | ⋮ | ⋮ | ⋱ | ⋮ |
| i | $N^1_1, i_1$ | $N^1_2, i_2$ | ... | $N^1_{g}, i_g$ |
| ⋮ | ⋮ | ⋮ | ⋱ | ⋮ |
| k | $N^k_1, k_1$ | $N^k_2, k_2$ | ... | $N^k_{g}, k_g$ |

**Table 2:** Pooled data passed from the nodes to the Analytical Center. We assume a 1:1 matched design. Here $V^1_{i(1)}$ ($V^2_{i(1)}$) denotes the pooled level of variable 1 (variable 2) for the i$^{th}$ case pool and $V^1_{i(0)}$ ($V^2_{i(0)}$) denotes the pooled level of variable 1 (variable 2) for the i$^{th}$ control pool, i= 1, 2, ..., k, where each pool consists of either *g* cases or *g* matched controls.

| Pool id | Pooled Covariate $V_1$ | | Pooled Covariate $V_2$ | |
|---|---|---|---|---|
| | Case | Control | Case | Control |
| 1 | $V^1_{1(1)}$ | $V^1_{1(0)}$ | $V^2_{1(1)}$ | $V^2_{1(0)}$ |
| ⋮ | ⋮ | ⋮ | ⋱ | ⋮ |
| i | $V^1_{i(1)}$ | $V^1_{i(0)}$ | $V^2_{i(1)}$ | $V^2_{i(0)}$ |
| ⋮ | ⋮ | ⋮ | ⋱ | ⋮ |
| k | $V^1_{k(1)}$ | $V^1_{k(0)}$ | $V^2_{k(1)}$ | $V^2_{k(0)}$ |

**Table 3:** Individual participant data for Toy example.

| Node | Stratum | Subject id | Outcome | Age | Gender | Marker | Pool id |
|---|---|---|---|---|---|---|---|
| Blue | 1 | 01 | 1 | 56 | 1 | 116 | 1 |
| Blue | 1 | 02 | 0 | 27 | 1 | 82 | 1 |
| Blue | 1 | 03 | 0 | 38 | 1 | 96 | 1 |
| Blue | 2 | 04 | 1 | 57 | 0 | 106 | 2 |
| Blue | 2 | 05 | 0 | 27 | 1 | 100 | 2 |
| Blue | 2 | 06 | 0 | 45 | 0 | 100 | 2 |
| Blue | 3 | 07 | 1 | 39 | 0 | 82 | 3 |
| Blue | 3 | 08 | 0 | 43 | 1 | 81 | 3 |
| Blue | 3 | 09 | 0 | 25 | 1 | 110 | 3 |
| Red | 4 | 10 | 1 | 49 | 0 | 103 | 1 |
| Red | 4 | 11 | 0 | 26 | 0 | 106 | 1 |
| Red | 4 | 12 | 0 | 43 | 0 | 111 | 1 |
| Red | 5 | 13 | 1 | 34 | 1 | 118 | 2 |
| Red | 5 | 14 | 0 | 36 | 0 | 93 | 2 |
| Red | 5 | 15 | 0 | 49 | 0 | 98 | 2 |
| Red | 6 | 16 | 1 | 34 | 1 | 93 | 2 |
| Red | 6 | 17 | 0 | 58 | 0 | 100 | 2 |
| Red | 6 | 18 | 0 | 33 | 1 | 90 | 2 |
| Green | 7 | 20 | 0 | 22 | 0 | 89 | 1 |
| Green | 7 | 21 | 0 | 45 | 1 | 100 | 1 |
| Green | 8 | 22 | 1 | 23 | 0 | 96 | 1 |
| Green | 8 | 23 | 0 | 56 | 1 | 97 | 3 |
| Green | 8 | 24 | 1 | 48 | 0 | 93 | 3 |
| Green | 9 | 25 | 0 | 24 | 0 | 106 | 3 |
| Green | 7 | 19 | 1 | 53 | 0 | 100 | 3 |
| Green | 9 | 26 | 0 | 50 | 0 | 109 | 3 |
| Green | 9 | 27 | 0 | 34 | 1 | 98 | 3 |

**Table 4:** Pooled data for Toy example. Data are fit such that the following model can be fit with the pooled levels.

$logit \Pr(D_{ij} = 1 | A_{ij} = a, G_{ij} = g, M_{ij} = m) = = \alpha_i + \beta \times \log a + \gamma \times g + \delta \times m + \vartheta \times \log a \times m.$

| Pool id | Outcome | Pooled levels | | | |
|---|---|---|---|---|---|
| | | log Age | Gender | Marker | log Age x marker |
| 1 | 1 | 11.01 | 1 | 308 | 1143.62 |
| 1 | 0 | 10.36 | 2 | 288 | 999.10 |
| 1 | 0 | 10.53 | 1 | 303 | 1069.84 |
| 2 | 1 | 11.10 | 2 | 317 | 1172.14 |
| 2 | 0 | 10.94 | 1 | 293 | 1066.34 |
| 2 | 0 | 11.19 | 1 | 288 | 1077.26 |
| 3 | 1 | 11.66 | 1 | 279 | 1089.14 |
| 3 | 0 | 11.54 | 1 | 283 | 1088.57 |
| 3 | 0 | 9.92 | 2 | 314 | 1039.25 |

**Table 5:** Parameter estimates between standard analysis and pooled analysis with different poolsizes for a binary outcome with a matched design. (see Results section for detailed simulation setting). In addition, the Monte Carlo Standard Error (MCSE), model-based SE (ModelSE) and coverage (nominal: 0:95) are also shown.

| Parameters | Unpooled | Pooled | | |
|---|---|---|---|---|
| | | $g = 4$ | $g = 6$ | $g = 10$ |
| $\beta = 0.3$ | | | | |
| Estimate | 0.301 | 0.303 | 0.306 | 0.331 |
| MCSE | 0.015 | 0.023 | 0.03 | 0.066 |
| ModelSE | 0.014 | 0.022 | 0.029 | 0.056 |
| Coverage | 0.942 | 0.942 | 0.946 | 0.958 |
| $\gamma = 0.2$ | | | | |
| Estimate | 0.202 | 0.203 | 0.200 | 0.225 |
| MCSE | 0.077 | 0.100 | 0.120 | 0.186 |
| ModelSE | 0.076 | 0.099 | 0.119 | 0.189 |
| Coverage | 0.948 | 0.944 | 0.958 | 0.978 |
| $\delta = 0.15$ | | | | |
| Estimate | 0.153 | 0.152 | 0.154 | 0.165 |
| MCSE | 0.039 | 0.049 | 0.060 | 0.100 |
| ModelSE | 0.037 | 0.049 | 0.059 | 0.094 |
| Coverage | 0.928 | 0.944 | 0.942 | 0.958 |
| $\omega = 0.09$ | | | | |
| Estimate | 0.093 | 0.092 | 0.094 | 0.098 |
| MCSE | 0.048 | 0.062 | 0.075 | 0.118 |
| ModelSE | 0.049 | 0.062 | 0.075 | 0.117 |
| Coverage | 0.962 | 0.956 | 0.950 | 0.962 |
| $\vartheta = 0.05$ | | | | |
| Estimate | 0.049 | 0.05 | 0.051 | 0.055 |
| MCSE | 0.014 | 0.017 | 0.023 | 0.038 |
| ModelSE | 0.013 | 0.018 | 0.022 | 0.036 |
| Coverage | 0.942 | 0.958 | 0.958 | 0.966 |

**Table 6:** Comparison between pooled analysis and unpooled analysis of matched case-control study on infertility. IA: any induced abortion, SA: any spontaneous abortion, EM: interaction between IA and SA. Dataset description: data.0,: data on 82 clusters; data.unpooled: data on 1000 clusters bootstrapped from data.0; data.pooled: pooled data using *g=2* from data.unpooled.

| Variable | OR (95% CI) | | |
|---|---|---|---|
| Dataset | data.0 | data.unpooled | data.pooled |
| IA | 5.27 (1.71, 16.27) | 6.91 (4.88, 9.78) | 5.76 (3.96, 8.37) |
| SA | 14.74 (4.77, 45.56) | 17.94 (12.69, 25.37) | 15.41 (10.40, 22.84) |
| IAxSA | 0.28 (0.06, 1.26) | 0.26 (0.17, 0.41) | 0.29 (0.18, 0.47) |

[Appendix]
[Include R code?]

References

1. Fears R, Brand H, Frackowiak R, Pastoret PP, Souhami R, Thompson B. Data protection regulation and the promotion of health research: getting the balance right. QJM. 2014 Jan;107(1):3-5. PubMed PMID: 24259722.
2. Homer N, Szelinger S, Redman M, Duggan D, Tembe W, Muehling J, et al. Resolving Individuals Contributing Trace Amounts of DNA to Highly Complex Mixtures Using High-Density SNP Genotyping Microarrays. PLoS Genet. 2008;4(8):e1000167 (1-9).
3. Singel R. Netflix Spilled Your Brokeback Mountain Secret, Lawsuit Claims 2009 [cited 2016 March 3]. Available from: http://www.wired.com/2009/12/netflix-privacy-lawsuit/.
4. Netflix Prize 2016 [cited 2016 March 3]. Available from: https://en.wikipedia.org/wiki/Netflix_Prize.
5. Rosano G, Pelliccia F, Gaudio C, Coats AJ. The challenge of performing effective medical research in the era of healthcare data protection. Int J Cardiol. 2014 Dec 15;177(2):510-1. PubMed PMID: 25183536.
6. Dolgin E. New data protection rules could harm research, science groups say. Nature Medicine. 2014;20(3):224.
7. Mostert M, Bredenoord AL, Biesaart MCIH, van Delden JJM. Big Data in medical research and EU data protection law: challenges to the consent or anonymise approach. Eur J Hum Genet. 2015 11/11/online.
8. Nyrén O, Stenbeck M, Grönberg H. The European Parliament proposal for the new EU General Data Protection Regulation may severely restrict European epidemiological research. European Journal of Epidemiology. 2014;29(4):227-30.
9. Olsen. Data protection and epidemiological research: a new EU regulation is in the pipeline. International Journal of Epidemiology. 2014;43(5):~ 1353-4.
10. Ploem MC, Essink-Bot ML, Stronks K. Proposed EU data protection regulation is a threat to medical research. BMJ. 2013 2013-05-31 14:56:20;346.
11. Vandenbroucke JP, Olsen J. Informed consent and the new EU regulation on data protection. International Journal of Epidemiology. 2013 December 1, 2013;42(6):1891-2.
12. Wartenberg D, Thompson WD. Privacy Versus Public Health: The Impact of Current Confidentiality Rules. American Journal of Public Health. 2010 2010/03/01;100(3):407-12. PubMed PMID: 20075316. Pubmed Central PMCID: 2820076.
13. Proposal for a Regulation of the European Parliament and of the Council on the protection of individuals with regard to the processing of personal data and on the free movement of such data (General Data Protection Regulation) [first reading], (2016).
14. Thompson B. Data protection: how medical researchers persuaded the European Parliament to compromise [Internet]2016. [2016-03-02]. Available from: http://blogs.lse.ac.uk/brexitvote/2016/02/11/data-protection-how-medical-researchers-persuaded-the-european-parliament-to-compromise/.
15. Saha-Chaudhuri P. Covariate Microaggregation for Logistic Regression: An Application for Analysis of Confidential Data.
16. Dorfman R. The Detection of Defective Members of Large Populations. Annals of Mathematical Statistics. 1943;14:436-40. PubMed PMID: WOS:000188517500039. English.
17. Kline RL, Brothers TA, Brookmeyer R, Zeger S, Quinn TC. Evaluation of human immunodeficiency virus seroprevalence in population surveys using pooled sera. Journal of Clinical Microbiology. 1989 July 1, 1989;27(7):1449-52.

18. Weinberg CR, Umbach DM. Using pooled exposure assessment to improve efficiency in case-control studies. Biometrics. 1999 Sep;55(3):718-26. PubMed PMID: WOS:000082683000007. English.
19. Saha-Chaudhuri P, Umbach DM, Weinberg CR. Pooled Exposure Assessment for Matched Case-Control Studies. Epidemiology. 2011 Sep;22(5):~704-12. PubMed PMID: 21747285. Pubmed Central PMCID: 3160274.
20. Lyles RH, Mitchell EM, Weinberg CR, Umbach DM, Schisterman EF. An Efficient Design Strategy for Logistic Regression Using Outcome and Covariate-Dependent Pooling of Biospecimens Prior to Assay. Biometrics. 2016.
21. Karr AF, Fulp WJ, Vera F, Young SS, Lin X, Reiter JP. Secure, privacy-preserving analysis of distributed databases. Technometrics. 2007;49:~335-45.
22. Yu OHY, Filion KB, Azoulay L, Patenaude V, Majdan A, Suissa S. Incretin-Based Drugs and the Risk of Congestive Heart Failure. Diabetes Care. 2015 Feb;38(2):~277-84. PubMed PMID: 25205143.
23. Herrett E, Gallagher AM, Bhaskaran K, Forbes H, Mathur R, van Staa T, et al. Data Resource Profile: Clinical Practice Research Datalink (CPRD). International Journal of Epidemiology. 2015.
24. Trichopoulos D, Handanos N, Danezis J, Kalandidi A, Kalapothaki V. Induced abortion and secondary infertility. Br J Obstet Gynaecol. 1976 Aug;83(8):645-50. PubMed PMID: 952796.
25. Wolfson M, Wallace SE, Masca N, Rowe G, Sheehan NA, Ferretti V, et al. DataSHIELD: resolving a conflict in contemporary bioscience-performing a pooled analysis of individual-level data without sharing the data. International Journal of Epidemiology. 2010;39:~1372-82.
26. Fienberg SE, Fulp WJ, Slavkovic AB, Wrobel TA. "Secure" Log-Linear and Logistic Regression Analysis of Distributed Databases. Privacy in Statistical Databases. 2006.
27. Domingo-Ferrer JM, Mateo-Sanz J. Practical data-oriented microaggregation for statistical disclosure control. IEEE Transactions on Knolwedge and Data Engeneering. 2002;14:~189-201.
28. Schmid M. Estimation of a linear model under microaggregation by individual ranking. Allgemeines Statistisches Archiv. 2006;90(3):~419-38.
29. Schmid M, Schneeweiss H. The effect of microaggregation by individual ranking on the estimation of moments Journal of Econometrics 2009;153(2):~174-82.